\documentclass[journal=jacsat,manuscript=article]{achemso}

\usepackage[version=3]{mhchem} 
\usepackage{graphicx}
\usepackage{dcolumn}
\usepackage{bm}
\usepackage{siunitx}
\usepackage{amsmath}
\usepackage[dvipsnames]{xcolor}

\author{Hanqing Liu}
\email{H.Liu-7@tudelft.nl}
\affiliation[Delft University of Technology]
{Department of Precision and Microsystems Engineering, Delft University of Technology, Lorentzweg 1, 2628 CD Delft, The Netherlands}

\author{Gabriele Baglioni}
\affiliation[Delft University of Technology]
{Kavli Institute of Nanoscience, Delft University of Technology, 2628 CJ Delft, The Netherlands}

\author{Carla Boix-Constant}
\affiliation[Universitat de Valencia]
{Instituto de Ciencia Molecular (ICMol), Universitat de Valencia, Paterna 46980, Spain}

\author{Herre S. J. van der Zant}
\affiliation[Delft University of Technology]
{Kavli Institute of Nanoscience, Delft University of Technology, 2628 CJ Delft, The Netherlands}
\author{Peter G. Steeneken}
\affiliation[Delft University of Technology]
{Department of Precision and Microsystems Engineering, Delft University of Technology, Lorentzweg 1, 2628 CD Delft, The Netherlands}
\alsoaffiliation[Delft University of Technology]
{Kavli Institute of Nanoscience, Delft University of Technology, 2628 CJ Delft, The Netherlands}
\author{Gerard J. Verbiest}
\email{G.J.Verbiest@tudelft.nl}
\affiliation[Delft University of Technology]
{Department of Precision and Microsystems Engineering, Delft University of Technology, Lorentzweg 1, 2628 CD Delft, The Netherlands}

\title[An \textsf{achemso} demo]
    {Enhanced sensitivity and tunability of thermomechanical resonance near the buckling bifurcation}

\abbreviations{IR,NMR,UV}

\begin{document}


\begin{abstract}
The high susceptibility of ultrathin two-dimensional (2D) material resonators to force and temperature makes them ideal systems for sensing applications and exploring thermomechanical coupling. Although the dynamics of these systems at high stress has been thoroughly investigated, their behaviour near the buckling transition has received less attention. Here, we demonstrate that the force sensitivity and frequency tunability of 2D material resonators are significantly enhanced near the buckling bifurcation. This bifurcation is triggered by compressive displacement that we induce via thermal expansion of the devices, while measuring their dynamics via an optomechanical technique. We understand the frequency tuning of the devices through a mechanical buckling model, which allows to extract the central deflection and boundary compressive displacement of the membrane. Surprisingly, we obtain a remarkable enhancement of up to 14$\times$ the vibration amplitude attributed to a very low stiffness of the membrane at the buckling transition, as well as a high frequency tunability by temperature of more than 4.02~$\%$~\si{K^{-1}}. The presented results provide insights into the effects of buckling on the dynamics of free-standing 2D materials and thereby open up opportunities for the realization of 2D resonant sensors with buckling-enhanced sensitivity.
\end{abstract}
\section{Keywords}
Nanomechanical resonator, buckling bifurcation, frequency tuning, vibration amplitude  

\section{Introduction}

A flat mechanical plate subjected to a sufficiently high in-plane compressive stress becomes unstable, as its out-of-plane stiffness gradually reduces to zero \cite{hu2015buckling}. When this happens, the plate experiences a buckling bifurcation. Even the slightest imperfection in the device, like a very small initial deformation, can determine whether the plate buckles up or downward. This high sensitivity to initial conditions offers exciting prospects, both for studying material properties \cite{lindahl2012determination, le2021thermal, chil2022buckling} and for realizing new sensing applications \cite{el2019high}. Therefore there has been a growing interest for buckling in nano-electromechanical systems (NEMS) and resonators such as phononic waveguides \cite{kim2021buckling}, carbon nanotubes \cite{rechnitz2022dc} and SiN$_x$ drumheads \cite{kanj2022ultra}, showing reversible control of signal transmission, high sensitive switching, as well as remarkable nonlinear effects and high tunability of resonance frequencies. These properties of buckled resonators make them very suitable for applications as actuators, sensors, and energy harvesters \cite{xia2020nonlinear, harne2013review}.

Nanomechanical resonators made of free-standing 2D materials are stiff within the plane, due to their high Young's modulus, but extremely flexible out-of-plane due to their atomic thickness\cite{xu2022nanomechanical, steeneken2021dynamics, liu2023nanomechanical}. As a result, free-standing 2D materials buckle at relatively low compressive stress values and thereby present an ideal platform for studying the buckling bifurcation in nanoscale systems. In fact, the buckling bifurcation provides a sensitive method to determine the bending rigidity of 2D materials \cite{lindahl2012determination, iguiniz2019revisiting}. However, most of the work on 2D NEMS resonators has focused on flat 2D mechanical resonators under tensile stress, because these can be more reproducibly fabricated \cite{kim2020impact,ferrari2021dissipation}. Moreover, the experimental detection of the buckling bifurcation in 2D NEMS remains difficult, as it requires a methodology to induce symmetric in-plane compression in suspended 2D materials while measuring their mechanical motion with high spatial resolution. 

In this work, we study the effect of the buckling bifurcation on the dynamics of optothermally driven nanomechanical resonators made of FePS$_3$ membranes. By varying temperature, the membranes expand, causing a compressive displacement that triggers the membranes to deflect out-of-plane. Interestingly, this buckling bifurcation does not only cause a large change in the temperature-dependent resonance frequency, but also gives rise to a significant enhancement of vibration amplitude of the resonators when driven on-resonance. To account for these observations and relate them to the device parameters, we fit a mechanical buckling model to the experiments that quantifies central membrane deflection and boundary compressive displacement of the membrane. Based on the model we attribute the force response to a significantly reduced out-of-plane stiffness at buckling transition. The large frequency tuning and high responsivity to forces of 2D resonators near the buckling bifurcation might be utilized to enhance sensitivity in future designs of 2D NEMS devices like microphones and pressure sensors.

\section{Fabrication and methodology}
\begin{table}
\caption{\label{tab:table1}
Characteristics of the fabricated devices including radius $R$, thickness $h$, Young's modulus $E$, Poisson ratio $\nu$, mass density $\rho$, initial resonance frequency $f_0(T_0)$, temperature at turning point $T_{t}$, resonance frequency at turning point $f_t$, central deflection without boundary displacement loading $z_{free}$, initial displacement $U_0$, and pre-strain $\epsilon_0$. D1 and D2 are from the same nanoflake (see Fig.~\ref{fig:1}a).}
\begin{tabular}{llllll}
  \hline\hline
 &$R$ (\si{\micro m})&$h$ (\si{nm})&$E$ (\si{GPa}) & $f_0(T_0)$ (\si{MHz}) &$T_{t}$ (\si{K}) \\
\hline 
D1 & 4 & 33.9 & 69.9  & 5.89 & 302.0  
 \\
D2 & 4 & 33.9 & 70.3  & 5.52 & 303.0
\\
D3 & 3 & 34.5& 93.1 & 11.18 & 307.5 
\\
  \hline \hline
\end{tabular}
\\ [0.5cm]

\begin{tabular}{lllll}
  \hline \hline
 & $f_t$ (\si{MHz}) &$z_{free}$ (\si{nm}) & $U_0$ (\si{nm}) &$\epsilon_0 \times 10^{-5}$   \\
\hline
D1 & 5.53  & 20.1 & 0.08 & 1.84 
 \\
D2 & 3.97 &  6.2 & 0.08 & 1.99
\\
D3 & 6.89 & 4.3 &  0.04 &   1.32
\\
  \hline\hline
\end{tabular}
\end{table}

\begin{figure*}
\includegraphics[width=1\linewidth,angle=0]{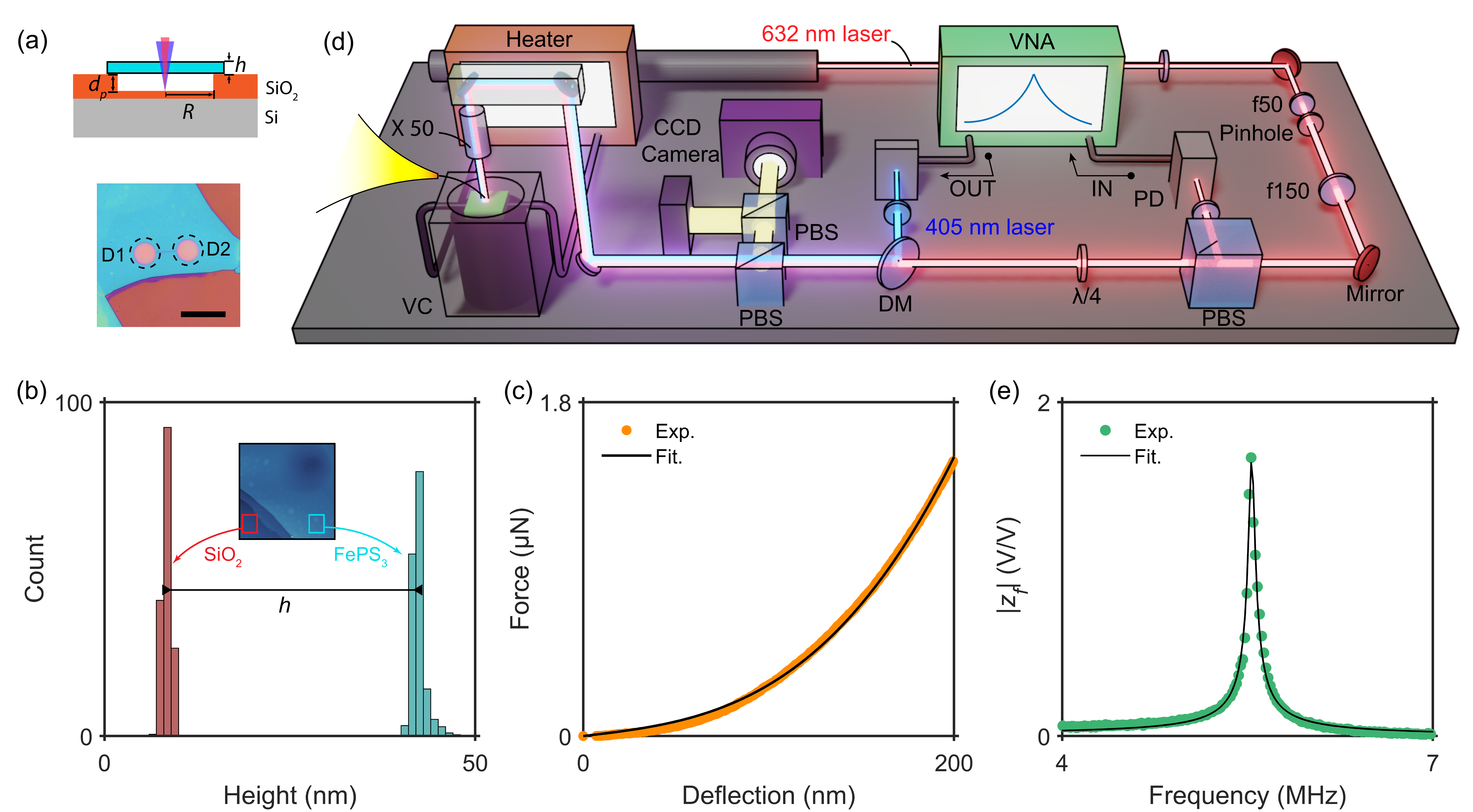}
\caption{\label{fig:1} {Sample characterization and optomechanical measurement.}  (a) Top, cross-section of a FePS$_3$ membrane suspended on the substrate with etched cavities. Bottom, optical images of the fabricated devices D1 and D2. Scale bar is 15~\si{\micro m}. (b) Height histogram of the substrate (red), as well as FePS$_3$ membrane (cyan), measured by AFM. Insert, AFM scanning image on the boundary of FePS$_3$ flake. (c) AFM indentation results for device D1 (orange points), where the Young's modulus $E$ of the membrane is extracted by fitting the measured force $F$ to the cantilever deflection $\delta$ (black line). (d) Intereferometry setup, where the chip is fixed inside a vacuum chamber (VC). VNA, vector network analyzer; PBS, polarized beam splitter; PD, photo diode; DM, dichroic mirror. (e) Measured signal $|z_f|$ around fundamental resonance mode (green points), which is fitted with a harmonic oscillator model to extract the resonance frequency $f_0$, quality factor $Q$ and the vibration amplitude $A_{\text{res}}$ (black line).} 
\end{figure*}

We fabricated 2D nanomechanical resonators by transferring exfoliated 2D flakes over etched circular cavities with a depth of $d_p=285$~\si{\nano\meter} and varying radius $R$ in a Si/SiO$_2$ substrate (Methods). In total, we made three FePS$_3$ devices D1$-$D3. Figure~\ref{fig:1}a shows a schematic cross-section and a top view (optical microscope) of the fabricated devices D1 and D2 with $R=4$~\si{\micro m}. Using tapping mode atomic force microscopy (AFM), we measure the height difference between the membrane and the Si/SiO$_2$ substrate. As Fig.~\ref{fig:1}b shows, we find a membrane thickness $h$ of 33.9~\si{nm} for devices D1 and D2. To determine the Young's modulus $E$ of the resonators, we indent the membrane centre by an AFM cantilever while measuring its deflection \cite{castellanos2012elastic}. We fit the applied force $F$ versus indentation $\delta$, as depicted in Fig.~\ref{fig:1}c (orange points), to a model for point-force loading of a circular plate given by $F = ( \frac{16\pi D}{R^2}\delta ) + n_0\pi\delta+Ehq^3(\frac{\delta^3}{R^2})$, where $D = Eh^3/(12(1-\nu^2))$ is the bending rigidity of the membrane, $\nu$ is Poisson ratio, $n_0=Eh\epsilon_0/(1-\nu)$ is the pre-tension in the membrane, and $\epsilon_0$ is the built-in strain. From the fit (black line, Fig.~\ref{fig:1}c) we extract $E=69.9$~\si{GPa} for device D1. The AFM measurements on devices D2 and D3 can be found in SI section 3. The extracted Young's moduli of all devices are listed in Table~\ref{tab:table1} and are similar to values reported in the literature \cite{vsivskins2020magnetic}.

To probe the thermodynamic properties of the fabricated devices, we use a laser interferometer (see Methods)\cite{Siskins2021tunable, liu2022tension}. As shown in Fig.~\ref{fig:1}d, we place the samples in a vacuum chamber with a pressure below 10$^{-5}$ \si{mbar} during the measurements. A power-modulated diode laser ($\lambda=405~$\si{nm}) photothermally actuates the resonator, while the reflection of a He-Ne laser ($\lambda=632$~\si{nm}) from the cavity with the suspended membrane captures its motion. The reflection is measured by a photodetector (PD) and processed by a Vector Network Analyzer (VNA) and then converted to the response amplitude $|z_f|$ of the resonator in the frequency domain. Figure~\ref{fig:1}e shows the measured frequency response around the fundamental resonance (green points) and a fit to a harmonic oscillator model (black line), given by $|z_{f}| = \frac{A_{\text{res}}f_0^2}{Q\sqrt{(f_0^2-f^2)^2+(f_0f/Q)^2}}$, where $f_0$ is resonance frequency, $A_{\text{res}}$ is the vibration amplitude at resonance and $Q$ is quality factor. Here, we extract $f_0=5.57$~\si{MHz}, $Q=195.93$ and $A_{\text{res}} = 1.64$~\si{V/V} for device D1. We will now outline how the characteristics of the resonance frequency ($f_0$ and $A_{\text{res}}$) can be used to provide information about the temperature dependent properties of 2D material resonators, in particular near the buckling bifurcation. 


\section{Results and discussion}
 
\begin{figure} 
\includegraphics[width=1\linewidth,angle=0]{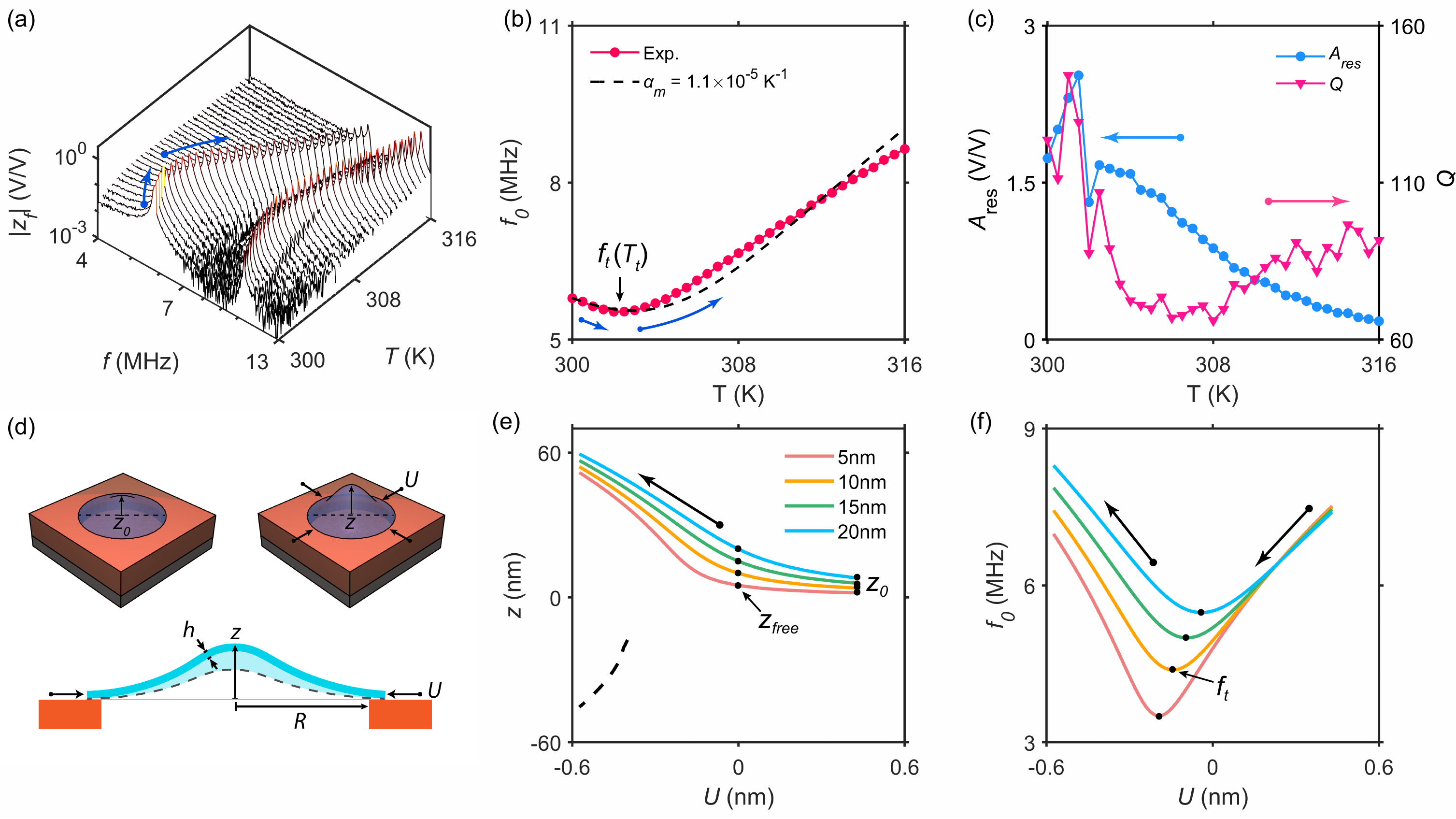}
\caption{{Thermally induced buckling in 2D nanomechanical resonator.} (a) Mechanical response $|z_f|$ of device D1 as the function of frequency $f$ and temperature $T$. The fundamental resonance frequency $f_0$ first decreases and then increases with increasing $T$ (blue arrows). (b) $f_0$ versus $T$ (red points). The minimum in $f_0$ is indicative of the temperature $T_t$ at around 302~\si{K}. Drawn line, $f_0$ versus $T$ fitted with the mechanical buckling model to the measurement, using $\alpha_m = 1.1\times 10^{-5}$~\si{K}$^{-1}$. The deviation at high $T$ indicates that $\alpha_m$ is $T$-dependent. (c) Vibration amplitude $A_{\text{res}}$ at resonance and quality factor $Q$ as a function of $T$ for device D1, respectively. (d) Mechanical buckling illustration for a clamped circular membrane, where a boundary compressive displacement $U$ causes a central deflection $z$ of the membrane. (e) $z$ versus $U$ in the membrane estimated by Eq.~\ref{eq.delfection}. Lines, results under different values of free deflection $z_{free}$ of the membrane. Dotted line, supercritical bifurcation at the critical buckling load when $z_{free}=5$~\si{nm}. (f) $f_0$ of the resonator versus $U$. Black dots, resonance frequency $f_t$ at the turning point. }
\label{Fig.buckling}
\end{figure}

Figure~\ref{Fig.buckling}a shows the measured $|z_f|$ as a function of actuation frequency and temperature (in the range from 300 to 316~\si{K}) for device D1. Interestingly, the resonance frequencies, including fundamental mode (indicated by the blue arrows) and second mode, first decrease and then increase as temperature increases, with a turning point at temperature $T_t = 302$~\si{K} (see Fig.~\ref{Fig.buckling}b). Similarly, the measured vibration amplitude $A_{\text{res}}$ for device D1 also reaches to its maximum at $T_t$ (see Fig.~\ref{Fig.buckling}c). These behaviors are also experimentally observed in devices D2 and D3 (see SI section 3). We attribute the turning of $f_0$ versus $T$ to the mechanical buckling of the nanomechanical resonators under critical compressive loading, which has been reported before in carbon nanotube resonators \cite{rechnitz2022dc} and arch MEMS devices \cite{hajjaj2020linear}. In fact, the bulk thermal expansion coefficients (TEC), $\alpha_m$, of the measured FePS$_3$ membranes in this work is much larger than the TEC $\alpha_\text{Si}$ of the Si/SiO$_2$ substrate. Hence, heating induces compressive displacement in the resonators and buckling is a natural consequence. 

Due to the buckling, we cannot use the standard equation for the resonance frequency of a pre-tensioned plate or membrane for further analysis. \textcolor{black}{Therefore, we use a mechanical buckling model for clamped circular plate, as illustrated in Fig.~\ref{Fig.buckling}d. Using a Galerkin method from literature \cite{yamaki1981non,kim1986flexural}, we obtain an expression of $f_0$ under thermally induced compressive displacement:} 
\begin{equation}
    f_0(T) =  \frac{10.33h}{\pi d^2}\sqrt{\frac{E}{3\rho (1-\nu^2)}(1 + \beta(1-\nu^2)\frac{3z^2-z_{free}^2}{h^2} + \frac{3}{8}(1+\nu)\frac{Ud}{h^2})},
    \label{eq.T-frequnecy}
\end{equation}
\textcolor{black}{where $U$ is the in-plane edge displacement of the plate, $\rho$ is the mass density, $z$ is the central deflection of the plate, $z_{free}$ is the central deflection of free plate without loading (when $U=0$), and $\beta $ is a fitting constant and equal to 0.52 when $\nu = 0.304$ (see SI section 1). Both $U$ and $z$ depend on temperature $T$. The details of derivation of Eq.~\ref{eq.T-frequnecy} can be found in SI section 1. In contrast to the standard equation for the resonance frequency of a pre-bending plate \cite{castellanos2013single}, we now find that not only the bending rigidity determine $f_0(T)$, but also the thermally-induced boundary displacement $U$ and the center deflection $z$ of the membrane. }

To find the relation between $U$ and $z$, we consider a uniformly-clamped plate as depicted in Fig.~\ref{Fig.buckling}d. \textcolor{black}{By studying the static state of the plate using Gakerkin method (see more details in SI section 1), we obtain an analytic solution:} 
\begin{equation}
   \frac{32}{3}\left(1-\frac{z_{free}}{z}\right) - 10.7\beta(1-\nu^2)\left(\frac{z_{free}^2-z^2}{h^2} \right) + 4(1+\nu)\frac{Ud}{h^2} = 0.
   \label{eq.delfection}
\end{equation}
\textcolor{black}{Therefore, the change of central deflection $z$ of the plate versus $U$ as buckling happens can be extracted from Eq.~\ref{eq.delfection}. We further use COMSOL simulation method to obtain $z$ and $f_0$ as the function of $U$, showing good agreements with the analytical solution obtained from Eqs.~\ref{eq.T-frequnecy} and \ref{eq.delfection} (see Fig.~S2).} By substituting the parameters $R=4$~\si{\micro \meter}, $E=69.9$~\si{GPa}, $h=33.9$~\si{\nano\meter}, $\rho=3.375$~\si{g/cm^3} and $\nu=0.304$ into Eq.~\ref{eq.delfection}, we can evaluate $z$ as a function of $U$ for different $z_{free}$. As plotted in Fig.~\ref{Fig.buckling}e, $z$ gradually increases with increasing $U$. The dotted black line in Fig.~\ref{Fig.buckling}e represents a supercritical bifurcation at the critical buckling load when $z_{free} = 5$~\si{nm}. This physically indicates an unstable equilibrium that the plate will either buckle up or down when it is slightly perturbed. For nonzero $z_{free}$ in this work, the 2D membrane always buckles in the direction of its pre-deflection.   

In order to investigate the effect of buckling on the resonance frequency $f_0$, we substitute the relation between $z$ and $U$ into Eq.~\ref{eq.T-frequnecy}. This results in a relation between $f_0$ and $z_{free}$, as plotted in Fig.~\ref{Fig.buckling}f. When decreasing the displacement $U$ by compression, $f_0$ reduces to a minimal value (the turning point) and then starts to increase. At this turning point, the minimum resonance frequency $f_t$ of the resonator is reached (marked as dots in Fig.~\ref{Fig.buckling}f). 
Both the experimental curves in Fig.~\ref{Fig.buckling}a and the theoretical curves in Fig.~\ref{Fig.buckling}f clearly show this frequency minimum, which we take as qualitative evidence for the occurrence of buckling in the 2D resonators.

\begin{figure} 
\includegraphics[width=1\linewidth,angle=0]{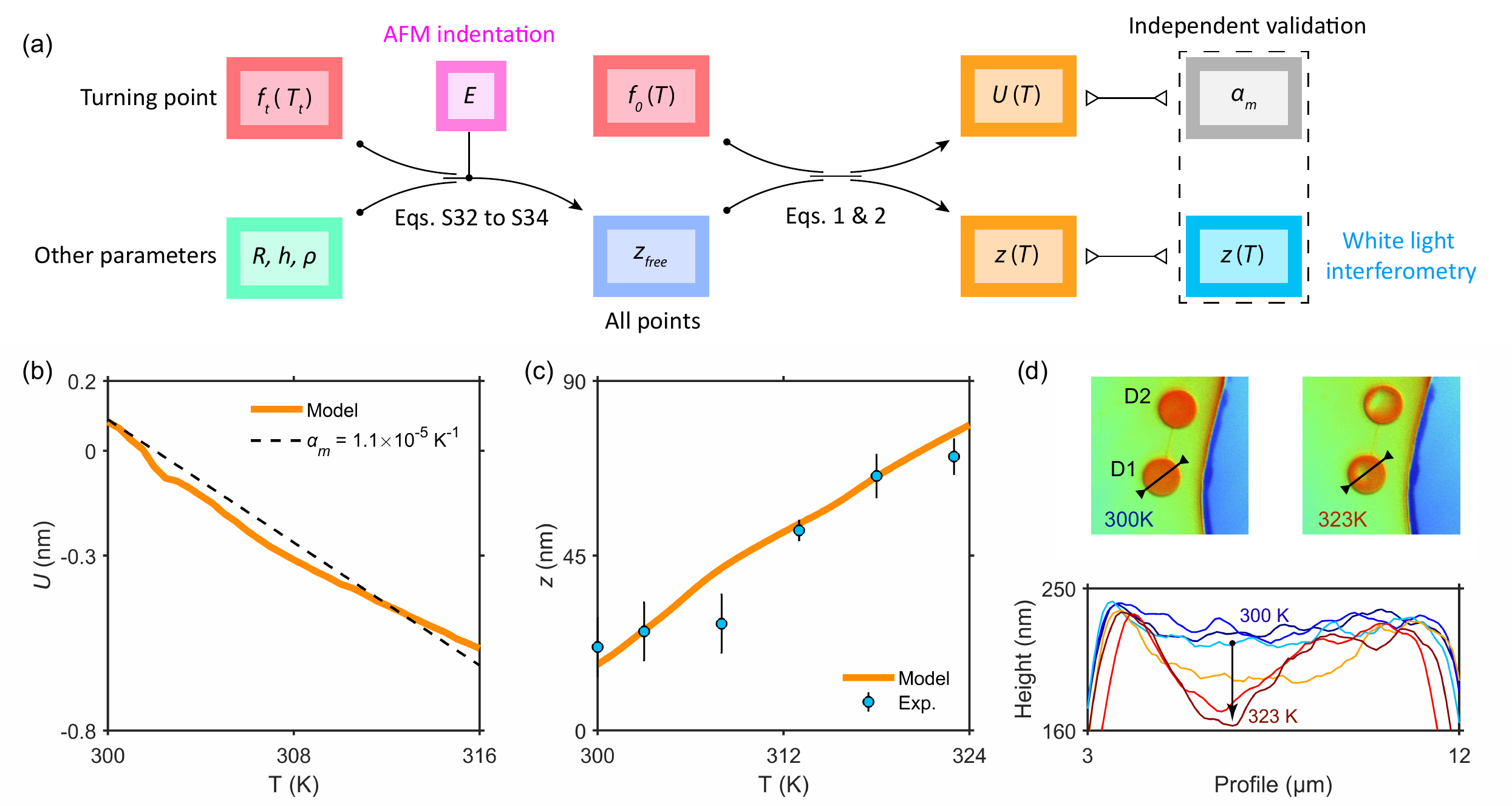}\hspace*{0cm}
\caption{{Change in boundary displacement and central deflection of device D1.} (a) Flow chart of the mechanical buckling model, which allows to extract the boundary displacement $U$ and central deflection $z$ of the membranes as the function of temperature $T$. (b) $U$ versus $T$. Orange line, the result obtained from mechanical buckling model; black lines, calculation with fixed values of TEC $\alpha_m$. (c) $z$ versus $T$. Orange line, the result obtained from mechanical buckling model; points, measurement by white light interferometry. (d) Top, images of FePS$_3$ flake under white light interferometry. Bottom, surface profile of the membrane as $T$ increases (corresponding to the black arrow in top insert). }
\label{Fig.experiment}
\end{figure}

\textcolor{black}{Let us now quantify $U$ and $z$ as a function of $T$ for device D1, using a model that follows the flow chart depicted in Fig.~\ref{Fig.experiment}a. First, we need to determine the value of $z_{free}$ in the mechanical buckling model. For this, we use the specific feature in the measured $f_0$ versus $T$ data, which is $\frac{\text{d} f_0}{\text{d} T}|_{T=T_{t}}=0$ at the turning point (Fig.~\ref{Fig.buckling}b). Assuming the Young's modulus of the membrane remains constant within the probed temperature range \cite{sha2015mechanical}, $z_{free}$ can be determined by $f_t$ (see derivation in SI section 2). Here, using the parameters in Table~\ref{tab:table1} and the measured value of $f_t(T_t) = 5.53$~\si{MHz}, we extract $z_{free} = 20.1$~\si{nm} for device D1.}


By substituting the obtained $z_{free}$ into Eqs.~\ref{eq.T-frequnecy} and \ref{eq.delfection}, we further extract $U$ and $z$ as a function of $T$ from the measured $f_0$ at each temperature for device D1. In Fig.~\ref{Fig.experiment}b, we observe that the compressive displacement $U$ becomes more than 10 times larger than its initial tensile value $U_0 = 0.08$~\si{nm} (Table~\ref{tab:table1}) by heating the membrane by only 16~\si{K}. \textcolor{black}{To validate the extracted $U(T)$, we determine the TEC $\alpha_m$ of the membrane. Using the TEC of the substrate $\alpha_{\text{Si}}$, we can use the relation $\frac{1}{R}\frac{\text{d} U}{\text{d} T} = -(\alpha_m-\alpha_{\text{Si}})$ to determine $\alpha_m$ \cite{vsivskins2020magnetic}.} We thus fit this relation to the obtained $\epsilon$ as shown in Fig.~\ref{Fig.experiment}c (orange line) and find $\alpha_m$ is approximately $1.1 \times 10^{-5}$~\si{K^{-1}}, which is in good agreement with values reported in the literature for FePS$_3$\cite{takano2004magnetic, vsivskins2020magnetic}. The fitting deviation for $f_0(T)$ in Fig.~\ref{Fig.buckling}b is thus attributed to the temperature dependence of $\alpha_m$ of FePS$_3$, or the irregular deflection of the membrane as buckling happens.       

To experimentally validate the extracted $z(T)$ from the buckling model in Fig.~\ref{Fig.experiment}c, we use a white light interferometer to image the surface profile of the suspended FePS$_3$ membranes as a function of temperature. As shown in Fig.~\ref{Fig.experiment}d, we observe from the height profiles (black arrows, top panel) that the membrane deformation increases as $T$ goes up (bottom panel). As a measure for $z(T)$, we take the difference between the maximum and minimum height for two height profiles and take the average value. As plotted in Fig.~\ref{Fig.experiment}c, the extracted $z(T)$ (points) for device D1 quantitatively matches the estimated $z(T)$ (orange line), which confirms that device D1 exhibits mechanical buckling. In addition, the total strain in the FePS$_3$ membrane also changes from the initial tensile strain to a strong compressive strain. From the obtained $U_0 = 0.08$~\si{nm} and $z_{free} = 20.1$~\si{nm}, we extract the initial strain $\epsilon_0 = 1.84\times 10^{-5}$ for device D1 using Eq.~S26 and the relation $N_r = \frac{Eh\epsilon}{1-\nu}$. All obtained $\epsilon_0$ for devices D1 to D3 are listed in Table~\ref{tab:table1}.

We now focus on the vibration amplitude of the fundamental mode of the membrane. As shown in Fig.~\ref{Fig.buckling}c, we observe a remarkable enhancement of up to $14\times$ the vibration amplitude $A_{\text{res}}$ at the turning point $T_t = 302$~\si{K}. This is attributed to the reduction of out-of-plane stiffness, $k_{\text{eff}} = m_{\text{eff}}(2\pi f_0)^2$, of the membrane near the buckling transition. Furthermore, we also find the thermally induced buckling in devices D2 and D3 during optomechanical measurements (see SI section 3). We quantify the frequency turning of these devices with the mechanical buckling model, and extract their $z_0$, $\epsilon_0$ and $\epsilon_t$ as listed in Table~1. Similar to what was observed for device D1 (Fig.~\ref{Fig.buckling}c), $A_{\text{res}}$ for devices D2 and D3 also show more than 14 times enhancements near the buckling transition (Fig.~S6). \textcolor{black}{However, although the quality factor $Q(T)$ reaches to its maximum at turning point for device D1, we observe completely different results of $Q(T)$ for devices D2 and D3. Since $Q$ is related to the TEC and thermal properties of 2D materials \cite{prabhakar2008theory}, it is of interest to investigate the $T$-tuning $Q$ of buckled 2D resonators in future work.} 
 
The implications of the observed phenomena extend beyond FePS$_3$ resonators. Even for 2D materials with a negative TEC such as graphene, buckling might occur if it is cooled down and the initial tensile stress is low enough. A key assumption in Eqs.~\ref{eq.T-frequnecy} and~\ref{eq.delfection} is a uniform compressive force at the boundary of the membrane and a constant Young's modulus over the measured temperature range. In reality, inhomogeneities due to uneven adhesion between membrane and substrate could lead to multiple smaller corrugations and wrinkles superimposed in the membrane when buckling occurs. This potential limitation, which we did not observe for the devices studied in this work, deserves future study as the buckled mode shape as well as the Young's modulus depend on it \cite{bonilla2016critical,nicholl2015effect}. Possibly, the experimental quantification of the Young's modulus for each device with AFM, as we did in Fig.~\ref{fig:1}c, compensates for some of the effects of corrugations and wrinkles on the buckling bifurcation.

Despite the fact that the temperature-dependence of resonance frequency has been investigated in earlier works on 2D membranes \cite{wang2021thermal, ye2018electrothermally, davidovikj2020ultrathin}, mechanical buckling has not been reported yet. It seems that one study on MoS$_2$ resonators might have almost reached the buckling point ($\frac{\text{d} f_0}{\text{d} T} \rightarrow 0$) at around 373~\si{K} \cite{wang2021thermal}. In this work it was relatively straightforward to reach the buckling bifurcation due to the large TECs of the selected 2D materials. It is of interest to speculate on the ultimate limits of buckling induced resonance frequency decrease. As indicated in Fig.~\ref{Fig.buckling}d, theoretically it might be possible to have the resonance frequency approaching zero for a deflection $z_{free}=0$~\si{nm}. However, in practice it will be difficult to reach that point. Nevertheless, by making the membranes flatter and with low pre-stress, the zero resonance frequency might be approached, which allows for an extremely high tunability and therefore a high force, stress and temperature sensitivity of $f_0$ near the minimum $f_t$. For such flat and low stress membranes, we expect the bending rigidity of 2D materials to dominate the performance and resonance frequency versus temperature curve near the buckling bifurcation point.

\begin{figure}
\includegraphics[width=8cm]{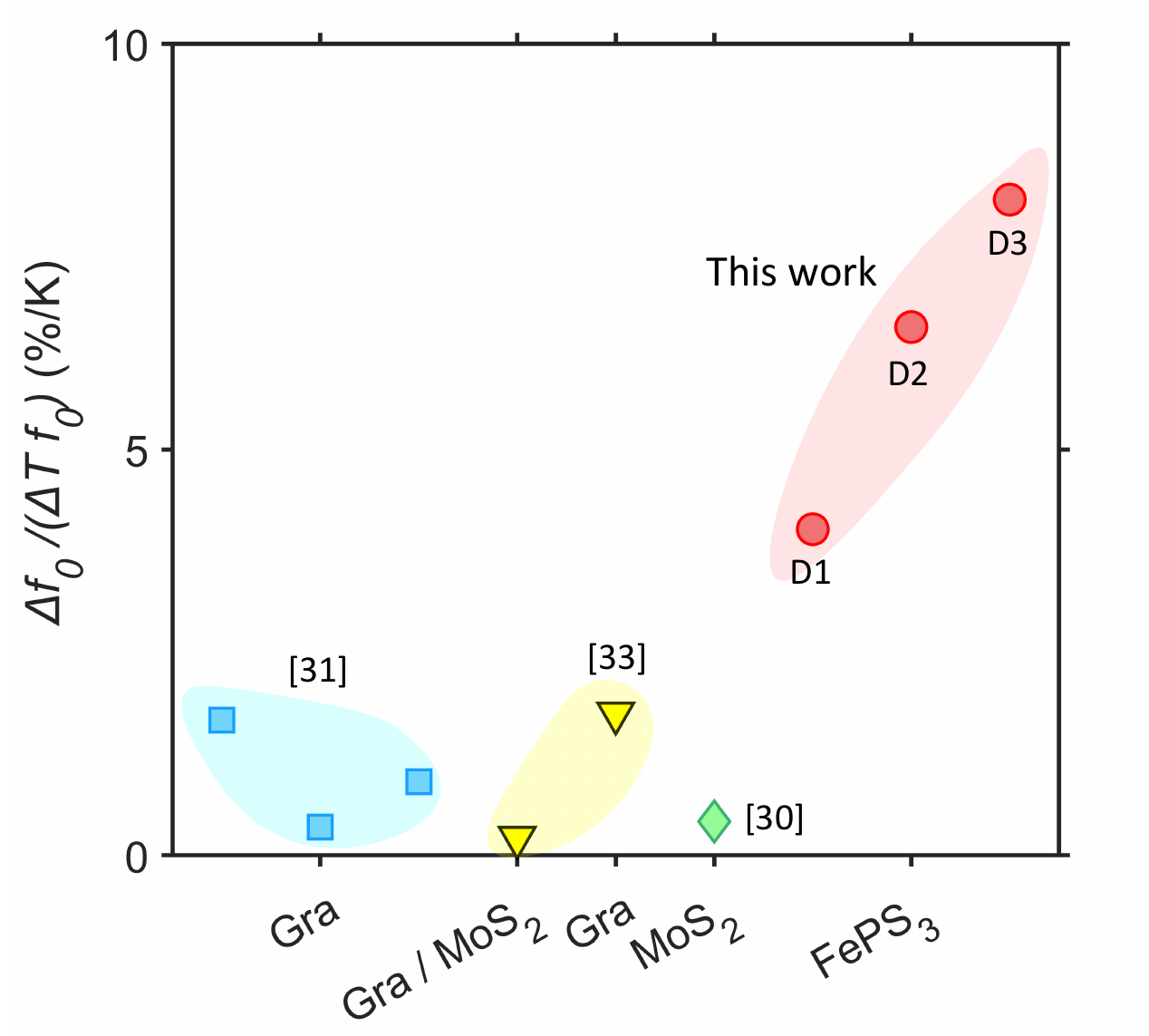}
\caption{ {Frequency tunability by varying temperature for 2D nanomechanical resonators}. }
\label{Fig.Thermal tunability}
\end{figure}

Also from an application perspective, the thermally induced buckling in 2D nanomechanical resonators deserves further exploration. First, the frequency tuning with temperature is considerable. As shown in Fig.~\ref{Fig.Thermal tunability}, we obtain a tunability, $\Delta f_0/(f_{t} \Delta T)$, more than 4.02~$\%$~\si{K^{-1}} for device D1, which is at least 2.3 times higher than reported in earlier studies \cite{ye2018electrothermally, inoue2018tuning, wang2021thermal}. The slope of frequency tuning for device D1 is $\Delta f_0/(\Delta T) = 194.3$~\si{kHz/K}, which, when considering an accuracy of $1$~\si{kHz} in determining $f_0$, results in a temperature resolution of $5.1$~\si{mK}. This value is comparable to state-of-the-art temperature sensors \cite{WinNT} and thus highlights the application buckled 2D resonators as bolometer \cite{blaikie2019fast} and NEMS resonant infrared detector \cite{qian2016graphene}. Furthermore, precise control over the buckling bifurcation can be obtained by tailoring the initial deflection of the membrane by applying, among others, electrostatic gating on the resonators \cite{steeneken2021dynamics}, a gas pressure difference \cite{sarafraz2022dynamics}, or straining the resonators by MEMS actuators \cite{verbiest2018detecting}.

\section{{Conclusion}}
In summary, we reported the experimental observation of thermally-induced buckling in 2D nanomechanical resonators made of suspended FePS$_3$ membranes. Using an optomechanical method, we probed their dynamic responses as a function of temperature. A mechanical buckling model was developed to explain the observed large turning of the resonance frequency with temperature, which allows to determine the boundary compressive displacement and center deflection of the fabricated devices. Using white light interferometer, we independently validated the extracted deflection of the membrane versus temperature from buckling model. We found an enhancement of up to 14$\times$ vibration amplitude near buckling bifurcation, which we attributed to the decrease in out-of-plane stiffness of the membrane. The gained insight not only advances the fundamental understanding of buckling bifurcation membranes made of 2D materials, but also enables pathways for buckling-enhanced designs and applications such as temperature detectors, thermoelectric and NEMS devices.

\section{{Methods}}
\textbf{Sample Fabrication.} A Si wafer with 285~\si{nm} dry SiO$_2$ is spin coated with positive e-beam resist and exposed by electron-beam lithography. Afterwards, the SiO$_2$ layer without protection is completely etched using CHF$_3$ and Ar plasma in a reactive ion etcher. The edges of cavities are examined to be well-defined by scanning electron microscopy (SEM) and AFM. After resist removal, FePS$_3$ nanoflakes are exfoliated by Scotch tape, and then separately transferred onto the substrate at room temperature through a deterministic dry stamping technique. \textcolor{black}{More details about the fabrication of etched substrate with circular cavities, as well as the Scotch tape transfer method can be found in \cite{liu2023nanomechanical}.} Detailed descriptions of the FePS$_3$ crystal growth and characterization is reported in earlier work \cite{ramos2021ultra}. \textcolor{black}{We choose FePS$_3$ flakes due to its large value of TEC, which allows to experimentally observe the buckling phenomenon of 2D resonators within a small range of temperature increase.} \\

\noindent\textbf{Laser Interferometry Setup.} We present temperature-dependent  optomechanical measurements in a laser interferometry setup \cite{vsivskins2020magnetic}. The fabricated devices is fixed on a sample holder inside the vacuum chamber. A PID heater and a temperature sensor are connected with the sample holder, which allows to precisely monitor and control the temperature sweeping. A piezo-electric actuator below the sample holder is used to optimize the X-Y position of the sample to maintain both the blue and red laser in the center of the 2D resonators. We use a red and blue laser power of 0.9 and 0.13~\si{mW} respectively. Note we verified that the resonators vibrate in linear regime and the temperature raise due to self-heating is negligible \cite{dolleman2018transient}. \textcolor{black}{The real picture of experimental setup on the optical table can be found in Fig.~S8. } 

\section{Associated content}
The Supporting Information is available free of charge at https://xxxxxx.

\section{Author information}
\subsection{Author Contribution}
H.L., P.G.S. and G.J.V. conceived the experiments. H.L. and G.B. performed the optomechanical measurements with heating control system. H.L. fabricated and inspected the samples. C.B.C. synthesized and characterized the FePS$_3$ crystals. H.L., G.J.V., P.G.S. and H.S.J.v.d.Z. analyzed and modeled the experimental data. H.S.J.v.d.Z. and P.G.S. supervised the project. The paper was jointly written by all authors with a main contribution from H.L. All authors discussed the results and commented on the paper.

\subsection{Notes}
The authors declare no competing financial interest.

\begin{acknowledgement}
P.G.S. and G.J.V. acknowledges support by the Dutch 4TU federation for the Plantenna project. H.L. acknowledges the financial support from China Scholarship Council. G.B., H.S.J.v.d.Z., and P.G.S. acknowledge funding from the European Union’s Horizon 2020 research and innovation
program under grant agreement no. 881603. C. B. C acknowledges the financial support from the European Union (ERC AdG Mol-2D 788222), the Spanish MICIN (2D-HETEROS PID2020-117152RB-100, co-financed by FEDER, and Excellence Unit “María de Maeztu” CEX2019-000919-M), the Generalitat Valenciana (PROMETEO Program and PO FEDER Program, Ph.D fellowship) and the Advanced Materials program (supported by MCIN with funding from European Union NextGenerationEU (PRTR-C17.I1) and by Generalitat Valenciana). We thank M.H. for checking the model derivations.  

\end{acknowledgement}



\bibliography{achemso-demo}

\end{document}